\documentclass[12pt,a4paper]{article}

\usepackage{enumitem}

\usepackage[T1]{fontenc} 
\usepackage{tgtermes} 

\usepackage{tabularray} 
\UseTblrLibrary{booktabs} 

\usepackage{graphicx} 
\usepackage[margin=25mm]{geometry}
\usepackage{amsmath}
\usepackage{amsfonts}
\usepackage{amssymb}
\usepackage{siunitx}
\usepackage{verbatim}
\usepackage{natbib} 
\usepackage{microtype} 

\usepackage[hyphens]{url}
\usepackage[hidelinks]{hyperref}
\usepackage[hyphenbreaks]{breakurl}

\title{Affirmative safety: \\
An approach to risk management for high-risk AI}
\author{
  Akash R. Wasil\\
  \small{Independent}\\
  \and 
  Joshua Clymer\\
  \small{Columbia University}\\
  \and
  David Krueger\\
  \small{University of Cambridge}\\
  \and
  Emily Dardaman\\
  \small{Independent}\\
  \and
  Simeon Campos\\
  \small{SaferAI}\\
  \and
  Evan R. Murphy\\
  \small{AI Security Initiative, Center for Long-Term Cybersecurity, UC Berkeley}\\
  \small{AI Governance \& Safety Canada}\\
}\date{}

\begin{document}
\maketitle

\begin{abstract}

\noindent
Prominent AI experts have suggested that companies developing high-risk AI systems should be required to show that such systems are safe before they can be developed or deployed. The goal of this paper is to expand on this idea and explore its implications for risk management. We argue that entities developing or deploying high-risk AI systems should be required to present evidence of “affirmative safety”– a proactive case that their activities keep risks below acceptable thresholds. We begin the paper by highlighting global security risks from AI that have been acknowledged by AI experts and world governments. Next, we briefly describe principles of risk management from other high-risk fields (e.g., nuclear safety). Then, we propose a risk management approach for advanced AI in which model developers must provide evidence that their activities keep certain risks below regulator-set thresholds. As a first step toward understanding what affirmative safety cases should include, we illustrate how certain kinds of technical evidence and operational evidence can support an affirmative safety case. In the technical section, we discuss behavioral evidence (evidence about model outputs), cognitive evidence (evidence about model internals), and developmental evidence (evidence about the training process). In the operational section, we offer examples of organizational practices that could contribute to affirmative safety cases: information security practices, safety culture, and emergency response capacity. Finally, we briefly compare our approach to the NIST AI Risk Management Framework. Overall, we hope our work contributes to ongoing discussions about national and global security risks posed by AI and regulatory approaches to address these risks. 

\end{abstract}
\newpage



\newpage
\section{Introduction}\label{sec:intro} 

\textbf{Advanced AI poses global security threats to humanity.} Leading AI researchers (such as Yoshua Bengio and Geoffrey Hinton) and the CEOs of three leading advanced AI companies (OpenAI, DeepMind, and Anthropic) have all signed a statement acknowledging: \textbf{“Mitigating the risk of extinction from AI should be a global priority alongside other societal-scale risks such as pandemics and nuclear war”} (Center for AI Safety, 2023). Sam Altman, CEO of OpenAI, stated that the bad case from AI is “lights out for all of us” (Loizos, 2023). Dario Amodei, CEO of Anthropic, claimed that the chance of a civilization-scale catastrophe resulting from AI was about 10-25\% (The Logan Bartlett Show, 2023). Geoffrey Hinton, considered a godfather of modern AI, recently quit Google to warn about the extinction risks from AI (MIT Technology Review, 2023). 

\textbf{World leaders have a responsibility to manage these risks.} Governments attending the UK AI Safety Summit acknowledged the potential for catastrophic harm in the Bletchley Declaration, a statement signed by representatives from all 28 participating countries (UK Government, 2023). They also recognized that frontier AI developers have a responsibility to ensure the safety of their systems (Prime Minister’s Office, 2023). The US AI Safety Institute, UK AI Safety Institute, the International Organization for Standardization, and the Cyberspace Administration of China have emerged to begin developing safety standards and evaluations that could become important building blocks for regulations (see China Law Translate, 2023; AI Safety Institute, 2023; National Institute of Standards and Technology, 2023b; ISO/IEC, 2023). 

\textbf{Safety standards for advanced AI should draw from best practices in risk management and emergency preparedness.} In high-risk industries with mature risk management infrastructure, the \textbf{burden of proof} is on the developer or manufacturer to show that their activities keep risks below an acceptable level. In this paper, we present the concept of “affirmative safety”, describe how it is applied in other fields, and offer suggestions about how to apply it to the regulation of advanced AI. 

\section{Principles of risk management}

\textbf{Risk management involves modeling a system’s potential outcomes, identifying vulnerabilities, and designing processes to define and reduce the likelihood of unacceptable outcomes.} In consumer goods manufacturing, such as infant cribs, regulators and/or stakeholders often require affirmative evidence of safety (CPSC, 2011). High-impact industries like nuclear energy, electrical engineering, and aviation have developed mature systems for risk prevention, monitoring, and response. 

\textbf{In high-stakes industries, risk management practices often require affirmative evidence that risks are kept below acceptable thresholds.} The Nuclear Regulatory Commission specifies that nuclear power plants must keep the risk of fatalities from reactor accidents below 0.1\%\footnote{The 0.1\% figure that the Nuclear Regulatory Commission (1983) employs for reactor accidents is specifically “0.1\% of certain other risks to which members of the U.S. population are exposed”, not a 0.1\% maximum probability over some span of time that such a fatal accident occurs.}, and reactor designs must show that the expected frequency of core damage is below 1 in 10,000 years (Nuclear Regulatory Commission, 1983; World Nuclear Association, 2022).  Similarly, the International Electrotechnical Commission requires qualitative or quantitative estimations of hazards for electronic safety-related systems. Failures that have a chance of occurring greater than 1/1000 per year are considered “frequent”, failures that occur between 1/10,000 and 1/100,000 are considered “occasional”, and failures that have a chance less than 1/10,000,000 are considered “incredible.” Consequences that result in multiple deaths are considered “catastrophic”, consequences that result in one death are considered “critical”, consequences that result in major injuries are considered “marginal”, and consequences that result in minor injuries are considered “negligible” (International Electrotechnical Commission, 2010).

\textbf{Demonstrating the absence of unacceptable risks requires more than simply showing that some risks have been addressed to some extent.} In areas where much is understood about a system, the standard is to show that risks are sufficiently rare in frequency and low in magnitude. A system’s risk of catastrophic failure is often a function of its complexity and how tightly coupled its components are (Perrow, 1999). This raises particular challenges for the effective governance of highly complex socio-technical systems like advanced AI models. 

\textbf{Robust risk management places an emphasis on \textit{understanding} how systems work.} If there is evidence that some aspect of a system is poorly understood, this is a source of risk (International Organization for Standardization, 2018). An understanding-based approach to risk management is especially important when trying to address events that are rare, difficult to predict, and have extremely large impacts on society (Nassim, 2007). As an illustration of this logic, Richard Feynman emphasized the importance of understanding failures for risk management when reflecting on the Challenger disaster:

\hfill\begin{minipage}{\dimexpr\textwidth-1cm}
“Erosion and blow-by are not what the design expected. They are \textbf{warnings that something is wrong.} The equipment is not operating as expected, and therefore there is a danger that it can operate with even wider deviations in this unexpected and not thoroughly understood way. The fact that this danger \textbf{did not lead to a catastrophe before} is \textbf{no guarantee that it will not the next time, unless it is completely understood.}” – Richard Feynman (reflecting on the Challenger disaster, emphasis added) (Feynman, 1986)
\end{minipage}
\hspace{1cm}

\section{A risk management approach for advanced AI}

In this section, we describe a risk management approach that could be applied to high-consequence AI systems\footnote{\textbf{Note on terminology.} The types of AI systems we are considering for our affirmative risk management approach are highly capable AI systems that present societal-scale risks. We use the terms “advanced AI”, “high-consequence AI” and “high-risk AI systems” more-or-less interchangeably to refer to these types of systems. Applying other terminology from the literature, we would include powerful forms of “general-purpose AI systems (GPAIS or GPAI; see Gutierrez et al. 2022). We would also include certain “frontier models”, defined as “cutting-edge, state-of-the-art, or highly capable GPAIS or foundation model[s]" (Barrett et al. 2023, p. 5). From a policy perspective, we would include "dual-use foundation models" as defined in the Biden Administration’s Executive Order: models that “pose a serious risk to security, national economic security, national public health or safety, or any combination of those matters.” This definition includes models that substantially lower the barrier of entry for designing or acquiring CBRN weapons, enabling powerful offensive cyberattacks, or permitting the evasion of human control or oversight."}.

Under this approach, \textbf{developers would be required to show regulators that they are keeping societal-scale risks below acceptable levels.} Regulators would be responsible for identifying \textbf{categories of risks} as well as \textbf{acceptable risk thresholds} for each category. For example, given that many AI experts are worried about risks from biological weapons from AI systems within the next 2-3 years (see Oversight of A.I.: Principles for Regulation, 2023), regulators might have “biological weapon development” as one of its risk categories. Given the extreme risks to public safety, regulators might set an acceptable risk threshold of 1/100,000 for this category: that is, an advanced AI developer must show that its development and deployment practices keep certain types of risks from AI-enabled biological weapons below 1/100,000. 

\textbf{Table 1 lists examples of potential risk categories and risk thresholds.}\footnote{Note: The risk categories and thresholds are included for illustration purposes. Our work is not intended to replace processes for developing alternative categories or thresholds.}

\begin{longtblr}[caption={Examples of potential risk categories and risk thresholds.},
  label={table:risk},
  ]{
    colspec={|Q[3cm,valign=m,halign=l]|X[l,valign=m]|X[l,valign=m]|},
    row{1}={font=\bfseries}, 
    rowhead=1,
  }
        \hline
        \textbf{Risk category} & \textbf{Description}  & \textbf{Example acceptable risk threshold} \\
        \hline 
        Biological weapons & AI-enabled biological weapons lead to a major global security risk & Highly unlikely (1/100,000)   \\
        \hline
        Concentration of power & AI systems lead to an unprecedented concentration of power without adequate societal precautions & Somewhat unlikely (1/1,000) \\
        \hline
        Cyberoffensive capabilities & AI-enabled cyberoffensive capabilities lead to a major national security risk & Highly unlikely (1/100,000) \\
        \hline
        Economic shock & AI-enabled automation leads to an unexpected economic shock without adequate societal preparations & Somewhat unlikely (1/1,000) \\
        \hline
        Increasing bias and discrimination & AI-enabled bias and discrimination leads to widespread, significant increases in discrimination in hiring, policing, or other meaningful sectors & Unlikely (1/10,000) \\  
        \hline 
        Major national security threat & AI systems pose a serious national security threat or threat to the continuity of government that is not adequately captured by other categories & Highly unlikely (1/100,000) \\
        \hline 
        Misinformation & AI-enabled misinformation leads to a major threat to national security or democratic institutions & Unlikely (1/10,000) \\
        \hline
        Widespread loss of control (WLC) & AI systems escape human control, potentially leading to human extinction or other catastrophic harms & Highly unlikely (1/100,000) \\
        \hline
\end{longtblr}

The risks listed in Table 1 are illustrative rather than exhaustive.  The risk thresholds are likewise illustrative and do not represent an overall prioritization of different issues; prioritization should consider existing harms as well as risks, and consider tractability, cost/benefit analysis, and ethicality of potential mitigations.  For alternative ways of classifying AI risks, we recommend readers to Kilian et al.’s AI Risk Classification framework, Hendrycks et al.’s overview of catastrophic risks, Critch and Russell’s TASRA, and OpenAI’s preparedness framework (Kilian et al., 2022; Hendycks et al., 2023; Critch \& Russell, 2023; OpenAI, 2024).

\section{Technical recommendations for affirmative safety}

We present three categories of evidence that regulators could use: \textbf{behavioral evidence} (evidence from model outputs), \textbf{cognitive evidence} (evidence from model internals), and \textbf{developmental evidence} (evidence from the training process).  

This taxonomy is meant to be a helpful heuristic for classifying various kinds of evidence, but the categories are neither mutually exclusive nor comprehensive. In each of these areas, there are already some promising ideas about the kind of evidence that could ensure that risks are below acceptable levels. However, new work will be needed, especially as AI systems become more powerful and more capable. For alternative and complementary perspectives, we recommend work on safety cases (Clymer et al., 2024), provably safe systems (Tegmark \& Omohundro, 2023), and safety for long-term planning agents (Cohen et al., 2024).

\subsection{Behavior: Robustly safe model outputs}
\textbf{Explanation: Regulators could require evidence that model behaviors are robustly safe and that models will act as intended.} 

One goal of AI safety research is to ensure that model outputs are safe and predictable across a wide array of possible inputs. For example, one may desire to build a model that cannot develop certain classes of biological weapons regardless of what prompt the model receives. Work on red-teaming and capabilities evaluations has focused on model outputs, attempting to identify if models are capable of dangerous outputs (e.g., OpenAI, 2023a). Broadly, evidence from an AI system’s behavior (outputs) becomes more compelling with the quantity, diversity, and representativeness of the input data points.  Adversarial testing (such as red-teaming) can also produce more compelling evidence. However, we note that current systems are highly vulnerable to adversarial attacks, and adversarial robustness may trade off against other forms of robustness. 

\textbf{Example: Testing generalization with “sandwiching” experiments.} Sandwiching experiments involve a novice, an AI system, and an expert. First, the AI is trained to perform a task and the novice provides human feedback to the AI system as part of its training. Then, the expert observes the trained system’s behavior and evaluates whether the AI system has learned the task correctly, or if it is simply providing incorrect behavior that the novice would perceive as correct. 

Human feedback is typically imperfect (see Christiano et al., 2017; Gao et al., 2022), and one of the primary goals of AI safety research is to ensure that models learn to adhere to human preferences despite imperfections in the oversight process. Sandwiching provides one way of evaluating the effectiveness of safety techniques: if we see that the AI system has learned to tell the novice what it thinks the novice “wants to hear”, we can conclude that the safety technique has not robustly prevented deception. For sufficiently advanced AI systems, it will be critical to have safety techniques that result in models that are truthful, as opposed to models that provide false-yet-believable answers. Sandwiching experiments are one tool we can use to examine how AI systems generalize in situations with imperfect oversight. 

\textbf{Related work:} Some researchers have conducted sandwiching experiments on frontier AI systems. For example, Anthropic researchers have examined sandwiching experiments in the context of scalable oversight techniques (Bowman et al., 2022). More recently, OpenAI researchers have developed a sandwiching setup to experiment with control and training techniques (OpenAI, 2023b). Specifically, they attempted to train GPT-4 to answer questions honestly by using a GPT-2 sized model for supervision. This is an analogy for training superhuman models with human oversight– in the analogy, GPT-2 is like the human overseer (a less intelligent agent providing supervision) and GPT-4 is like the superhuman model (a more intelligent agent being trained). Other researchers have proposed a broader generalization benchmark meant to test whether developers can control how honesty generalizes across a wide variety of distribution shifts (Clymer et al., 2023).

\subsection{Cognition: Understanding AI system internals}

\subsubsection{Empirical evidence on model internals}

\textbf{Explanation: Regulators could require empirical evidence that developers understand how their systems operate.} This includes showing that interpretability techniques and evaluations are robust to deception and misinterpretation. This could be evaluated using \textit{simulatability tests} that assess if developers can successfully predict the systems’ behavior in new situations.

Advanced AI systems may \textit{appear} trustworthy and helpful regardless of whether they actually are. One critical challenge of evaluating safety by observing external behavior is that AI systems might try to \textit{conceal} their motives (Park et al., 2023). If AI developers were able to reliably and robustly understand a model’s internal reasoning and show that this internal reasoning is harmless, this would provide evidence supporting the idea that they can keep risks below acceptable levels. 

\textbf{Example: Identifying deceptive models with interpretability tools.} Suppose two models have the same external behavior (outputs) across a set of circumstances, but they have been trained to engage in this behavior for \textit{different reasons} (different internal cognition). For example, one model might be trained to follow developer instructions, and another model could be (unintentionally) trained to follow instructions when monitored but \textit{disregard the instructions} if it infers that the evaluator is being inattentive. We consider the latter model to be "deceptively aligned" (see Park et al., 2023). 

Developers could evaluate whether available internal reasoning tools or interpretability tools can reliably distinguish between deceptive models and non-deceptive models. To pass this test, developers would need to show that they can distinguish between the models even when restricted to cases where their behaviors or outputs are identical. In other words, since the behavior of the two models would be identical (on the particular inputs selected), they would have to rely solely on their internal reasoning tools to distinguish between the two models. In practice, instead of just distinguishing between two models, there could be 100 models, with some unknown number of deceptive models. Additionally, the testers would run these tests multiple times in multiple different settings and environments to detect multiple different kinds of potentially dangerous qualities.

\textbf{Related work:} Much empirical work on understanding model internals has come from the field of interpretability (see Doshi-Velez \& Kim, 2017). When examining frontier AI systems, mechanistic interpretability research– a subset of interpretability research that focuses on understanding neural activations– has been especially common. Within mechanistic interpretability research, much work focuses on two high-level challenges: interpreting activations and understanding how the activations connect together to implement an algorithm (finding ‘circuits’). Researchers have identified interpretable concepts in groups of activations (Wang et al., 2022; Zou et al., 2023), devised methods for making individual activations more interpretable (Bricken et al., 2023), and automatically searched for interpretations of activations (Bills et al., 2023). Despite this progress, mechanistic interpretability is a young field, and there are not yet examples in which mechanistic interpretability research has yielded findings that could meaningfully enhance an affirmative argument for safety. Future work could aim to extend these approaches to identify concepts like “bioweapons” or “fraud” in AI activations. 

\subsubsection{Theoretical arguments about model internals}

\textbf{Explanation: Regulators could require formal and verifiable arguments that show that developers understand system internals.}

While empirical evidence is valuable, model internals are highly complex and may be difficult to make arguments about. Formal arguments that can be mathematically or logically verified offer an alternative approach. Even as AI systems become more advanced and formal arguments are too complicated for developers to understand, such arguments can still be verified. Leveraging formal arguments requires two steps: (1) finding a statement which, if true, would provide evidence for an AI system’s safety and (2) generating a proof of that statement.

\textbf{Example: Eliciting latent knowledge.} If developers could reliably determine what AI systems ‘believe,’ it would be much easier to trust and control them. However, for sufficiently powerful models, human overseers may not be able to trust the outputs of an AI system. There is already evidence that AI systems can sometimes recognize when they are in testing environments (Albert, 2024). Experts believe powerful systems will be even better at identifying when they are in test environments, rendering the results of those empirical safety tests unreliable (Cohen et al., 2024). In the eliciting latent knowledge report, researchers try to examine if there are strategies that could guarantee that models reveal their true beliefs (Christiano et al., 2021). However, there are currently no known ways to guarantee that powerful models report their beliefs accurately. Some researchers have attempted to develop a system for making formal statements about model ‘beliefs’’ (Christiano et al., 2022). While this approach is potentially promising, this research is in extremely early stages, and is not yet ready to be applied.

\textbf{Related work:} Formal verification of model behavior is an active ML research topic. The most common subproblem is ‘certified robustness’ – the problem of proving that a model’s output will not change if inputs are perturbed by some small amount (Li et al., 2023). So far, there are few examples of formal guarantees of safety properties for frontier AI systems.

\subsection{Development: Safe-by-design systems}

\textbf{Explanation: Regulators can require developers to provide formal verifications that powerful AI systems will behave safely within provable capability bounds.}

There has been great interest in safe-by-design AI systems: systems that have formal guarantees based on mathematical or logical proofs. Whereas the previous section focused on formal arguments that can be used to assess model internals after a model has already been developed, safe-by-design principles can be applied throughout the development cycle, ensuring that a resulting model is guaranteed to possess certain safety-relevant properties. For example, proofs could be applied to model architectures (e.g., to show that a certain training process has guaranteeable safety properties), hardware (e.g., to show that hardware provably meets certain security requirements), code (e.g., to show that code meets certain criteria that suggests that it can be run safely even if it is not fully understood), and various other steps (see Tegmark \& Omohundro, 2023). 

\textbf{Example: Researchers develop a new paradigm with theoretical and mathematical guarantees.} This paradigm allows for the safe development of highly powerful AI systems. Ideally, this paradigm would be competitive with deep learning (i.e., it would allow us to cost-effectively build powerful models just as well as deep learning or other state-of-the-art methods). More realistically, the safe-by-design paradigm may require more resources or time than alternative approaches. If such a safe-by-design paradigm were discovered, government intervention may be needed to ensure that models past a certain capabilities threshold could only be developed using safe-by-design architectures. 

\textbf{Related work:} Some researchers are investigating safe-by-design architectures that could scale toward artificial general intelligence. Examples include approaches focused on mathematical proofs that guarantee that models behave within certain quantitative bounds (e.g., Dalrymple, 2023; Dalrymple, 2024) and approaches focused on proof-carrying code (Tegmark \& Omohundro, 2023). Proof-carrying code could lead to automated software verification: mathematical proofs could be applied to code to guarantee that the code meets certain desired specifications. This approach could be necessary to verify that AI-generated code is safe to execute (see Tegmark \& Omohundro, 2023). 

\section{Operational practices for affirmative safety}

While our focus in this paper is on describing the technical components of affirmative safety, it is important to recognize that \textbf{operational practices} also play an important role. By “operational practices”, we refer to aspects of an organization’s culture, decision-making processes, and internal governance mechanisms that may increase or decrease certain kinds of risks. An exhaustive or comprehensive list of operational factors is outside the scope of this paper; we focus on three particularly important operational factors: \textbf{information security practices}, \textbf{safety culture}, and \textbf{emergency response capacity}.

\textbf{Information security.} Poor information security could lead to malicious actors stealing the weights of powerful AI systems. As a result, to show that an organization is keeping risks below acceptable risk thresholds, they may need to show that they have sufficient safeguards in place to protect sensitive material that could allow malicious actors to create dangerous AI systems (e.g., model weights). This principle is already present in Anthropic’s Responsible Scaling Policy: Anthropic publicly committed to not develop “ASL-3 systems” (AI that could substantially increase the risk of catastrophic misuse, for example by enabling large-scale biological attacks) until its information security standards were sufficiently strong “such that non-state attackers are unlikely to be able to steal model weights and advanced threat actors (e.g., states) cannot steal them without significant expense” (Anthropic, 2023). Ideally, rigorous information security standards would be established and checked by appropriate government officials and enforced across the board. 

\textbf{Safety culture.} Safety culture is commonly assessed in the realm of nuclear security. Broadly, safety culture refers to organizational practices that relate to how an institution and its members deal with concerns around safety and security. More formally, the International Atomic Energy Agency (IAEA) defines a strong safety culture as the “assembly of characteristics, attitudes, and behaviors in individuals, organizations and institutions which establishes that, as an overriding priority, protection and safety issues receive the attention warranted by their significance” (IAEA, 2016). The IAEA conducts safety culture assessments to review the culture of nuclear facilities and identify potential improvements. Operational Safety Review Teams (OSART), consisting of international experts with experience in nuclear safety, conduct these assessments. The assessments include on-site evaluations (observations of operating procedures, review of relevant documents), interviews and surveys with staff, and an examination of the organization's decision-making track record (for details, see IAEA, 2016). This process is used to assess several aspects of safety culture; examples include leadership's commitment to safety, safety training, communication processes, risk management procedures, attitudes toward safety, risk reporting systems, employee understanding of risks, and allocation of resources for safety. An affirmative case for safety could require organizations to provide evidence of their safety culture or receive sufficiently high scores on safety culture assessments conducted by independent parties.

\textbf{Emergency response capacity.} Risks from advanced AI might arise suddenly and with short notice. As a result, an affirmative safety case may require institutions developing advanced AI to show that they have sufficient measures in place to detect and manage sudden risks. This principle is present in OpenAI’s preparedness framework (OpenAI, 2024): OpenAI safety researchers can “fast-track” information to leadership if “a severe risk rapidly develops” (OpenAI, 2024). To expand on this, governments could require advanced AI companies to have emergency response plans that notify not only senior leadership at the AI company but also relevant national security figures or AI experts in the US government. In the event of an imminent AI-related emergency, it would be essential for government officials to be notified and have the ability to intervene. Emergency response plans must also dictate adequate responses that AI organizations and/or external parties could take if such a risk emerges. As an example, emergency response plans could involve protocols to implement “kill switches” that allow governments to swiftly halt a dangerous AI experiment or have a company safely withdraw access to a dangerous AI model (Miotti \& Wasil, 2023; Wasil, 2023).

\section{Comparison to NIST AI Risk Management Framework}

The National Institute of Standards and Technology (NIST) released the Artificial Intelligence Risk Management Framework (AI RMF). The framework describes desired criteria for AI systems: they ought to be (a) \textbf{valid and reliable}, (b) \textbf{safe}, (c) \textbf{fair and unbiased}, (d) \textbf{secure and resilient}, (e) \textbf{transparent and accountable}, (f) \textbf{explainable and interpretable}, and (g) \textbf{privacy-enhanced} (NIST, 2023a). NIST’s work is intended to offer a framework that can help companies reason about risks and make voluntary commitments.

NIST’s work differs from our recommendations in a few important ways. First, NIST’s Risk Management Framework is entirely voluntary – companies are free to ignore its recommendations. NIST describes the Risk Management Framework as “regulation-agnostic” and notes that the framework is not meant to supersede regulations and laws (NIST, 2023a). Second, and relatedly, NIST does not assign risk tolerance– it does not specify the level of risk that is considered acceptable in various domains. NIST’s work has valuably helped to introduce a common language when discussing risk management, to define desired criteria, and to pave the way for voluntary commitments. However, NIST recognizes the limitations of voluntary approaches, and the NIST framework should not be a substitute for binding regulations (NIST, 2023a). 

Notably, the NIST AI Risk Management Framework does recognize that certain kinds of AI development could pose unacceptably high-risk levels. “In cases where an AI system presents unacceptable negative risk levels – such as where significant negative impacts are imminent, severe harms are actually occurring, or catastrophic risks are present – development and deployment should cease in a safe manner until risks can be sufficiently managed” (NIST, 2023a). We agree strongly with this principle. Ideally, this principle would be instantiated by a regulatory body that reviews technical evidence (such as the evidence described above) and non-technical evidence (such as an organization’s safety culture and information security practices) to determine if risks are being sufficiently managed.

Finally, our affirmative AI safety approach is only intended for high-risk, advanced AI systems. Meanwhile, the NIST AI RMF was created to apply very generally to many different types of AI systems, including smaller models and lower-impact systems (for an application of the NIST AI RMF to general-purpose AI systems, see Barrett et al., 2023).

\section{Conclusion}

As AI systems continue to increase in their capabilities and their societal impacts, it will be important for regulators and policymakers to implement governance frameworks and standards that reduce risks. This will be especially important in the case of high-risk systems that are implemented in sensitive domains, general-purpose AI systems that can be used across a variety of domains, and frontier AI systems that could possess dangerous capabilities.

We described \textbf{affirmative safety} – the principle that risk management for advanced AI should go beyond a mere absence of clear evidence of danger. Instead, regulators could demand proactive or “affirmative” evidence that model developers understand how systems work and have adequate safeguards to keep certain risks below acceptable levels. To support our conceptual approach, we described specific technical evaluations and operational practices that could be used to support affirmative safety cases. We hope such work will be useful as legislative bodies, standards-setting organizations, and regulators develop risk management approaches for advanced AI systems. 

\section{Acknowledgments}
We extend our thanks to Callum Hinchcliffe for providing research assistance on this paper. We also would like to thank the following people who greatly improved this paper by providing comments on earlier drafts before publication: Alan Chan, Andrew McKnight, Michael Aird, and Yolanda Lannquist.



\pagebreak

\bibliographystyle{apalike}
\bibliography{example}
\sloppy

AI Safety Institute. (2023). \textit{Introducing the AI Safety Institute} \url{https://www.gov.uk/government/publications/ai-safety-institute-overview/introducing-the-ai-safety-institute}

Albert, A. [@alexalbert\_] (2024). “Fun story from our internal testing on Claude 3 Opus. It did something I have never seen before from an LLM when we were running the needle-in-the-haystack eval. [...]” X/Twitter. \url{https://twitter.com/alexalbert__/status/1764722513014329620}

Anthropic. (2023, September). \textit{Anthropic's Responsible Scaling Policy} \url{https://www-cdn.anthropic.com/files/4zrzovbb/website/1adf000c8f675958c2ee23805d91aaade1cd4613.pdf}

Barrett, A. M., Newman, J., Nonnecke, B., Hendrycks, D., Murphy, E. R., \& Jackson, K. (2023). \textit{AI risk-management standards profile for general-purpose AI systems (GPAIS) and foundation models.} Center for Long-Term Cybersecurity, UC Berkeley. \url{https://cltc.berkeley.edu/wp-content/uploads/2023/11/Berkeley-GPAIS-Foundation-Model-Risk-Management-Standards-Profile-v1.0.pdf}

Bills, S., Cammarata, N., Mossing, D., Tillman, H., Gao, L., Goh, G., ... \& Saunders, W. (2023). \textit{Language models can explain neurons in language models.} \url{https://openaipublic.blob.core.windows.net/neuron-explainer/paper/index.html}

Bricken, T., Templeton, A., Batson, J., Chen, B., Jermyn, A., Conerly, T., ... \& Olah, C. (2023). \textit{Towards monosemanticity: Decomposing language models with dictionary learning.} Transformer Circuits Thread, 2. \url{https://transformer-circuits.pub/2023/monosemantic-features}

Center for AI Safety. (2023). \textit{Statement on AI Risk.} \url{https://www.safe.ai/statement-on-ai-risk}

China Law Translate (2023). \textit{Interim Measures for the Management of Generative Artificial Intelligence Services.} \url{https://www.chinalawtranslate.com/en/generative-ai-interim/}

Christiano, P. F., Leike, J., Brown, T., Martic, M., Legg, S., \& Amodei, D. (2017). \textit{Deep reinforcement learning from human preferences.} \url{https://doi.org/10.48550/arXiv.1706.03741}

Christiano, P., Cotra, A., \& Xu, M. (2021). \textit{Eliciting latent knowledge: How to tell if your eyes deceive you.} \url{https://docs.google.com/document/d/1WwsnJQstPq91_Yh-Ch2XRL8H_EpsnjrC1dwZXR37PC8/}

Christiano, P., Neyman, E., \& Xu, M. (2022). \textit{Formalizing the presumption of independence.} \url{https://doi.org/10.48550/arXiv.2211.06738}

Clymer, J., Baker, G., Subramani, R., \& Wang, S. (2023). \textit{Generalization Analogies: A Testbed for Generalizing AI Oversight to Hard-To-Measure Domains.} \url{https://doi.org/10.48550/arXiv.2311.07723}

Clymer, J., Gabrieli, N., Krueger, D., \& Larsen, T. (2024). \textit{Safety Cases: Justifying the Safety of Advanced AI Systems.} 
\url{https://arxiv.org/abs/2403.10462}

Cohen, M. K., Kolt, N., Bengio, Y., Hadfield, G. K., \& Russell, S. (2024). \textit{Regulating advanced artificial agents.} Science, 384(6691), 36-38.

Critch, A., \& Russell, S. (2023). \textit{TASRA: A taxonomy and analysis of societal-scale risks from AI}. arXiv preprint \url{https://doi.org/10.48550/arXiv.2306.06924}

CPSC. (2011). \textit{CPSC Sets Crib Safety Standards.} \url{https://www.cpsc.gov/Newsroom/News-Releases/2012/CPSC-Sets-Crib-Safety-Standards}

Dalrymple, D. (2023). \textit{Mathematics and modelling are the keys we need to safely unlock transformative AI.} \url{https://www.aria.org.uk/wp-content/uploads/2023/10/ARIA-Mathematics-and-modelling-are-the-keys-we-need-to-safely-unlock-transformative-AI-v01.pdf}

Dalrymple, D. (2024). \textit{Safeguarded AI: constructing safety by design Programme thesis v1.0}
\url{https://www.aria.org.uk/wp-content/uploads/2024/01/ARIA-Safeguarded-AI-Programme-Thesis-V1.pdf}

Doshi-Velez, F., \& Kim, B. (2017). \textit{Towards a rigorous science of interpretable machine learning.} arXiv preprint arXiv:1702.08608

Feynman, R. P. (1986). Volume 2: Appendix F - Personal Observations on Reliability of Shuttle. \textit{Report of the Presidential Commission on the Space Shuttle Challenger Accident.} NASA. \url{https://www.nasa.gov/history/rogersrep/v2appf.htm}

Gao, L., Schulman, J. \& Hilton, J. (2022). \textit{Scaling Laws for Reward Model Overoptimization.} \url{https://doi.org/10.48550/arXiv.2210.10760}

Gutierrez, C. I., Aguirre, A., Uuk, R., Boine, C. C., \& Franklin, M. (2022). \textit{A proposal for a definition of general purpose artificial intelligence systems.} \url{https://dx.doi.org/10.2139/ssrn.4238951}

Hendrycks, D., Mazeika, M., \& Woodside, T. (2023). \textit{An overview of catastrophic AI risks.} arXiv preprint. \url{https://arxiv.org/abs/2306.12001}

International Atomic Energy Agency (IAEA) (2016). \textit{Performing Safety Culture Self-Assessments.} Safety Reports Series No. 83. Vienna. \url{https://www-pub.iaea.org/MTCD/Publications/PDF/Pub1682_web.pdf}

International Electrotechnical Commission. (2010). IEC 61508-1:2010  \url{https://webstore.iec.ch/publication/5515}

International Organization for Standardization. (2018). ISO 31000: Risk management - Guidelines (ISO 31000:2018). \url{https://www.iso.org/standard/65694.html}

Kilian, K. A., Ventura, C. J., \& Bailey, M. M. (2022). \textit{Examining the differential risk from high-level artificial intelligence and the question of control.} \url{https://arxiv.org/pdf/2211.03157.pdf}

Kosoy, V. (2023). \textit{The Learning-Theoretic Agenda: Status 2023} \url{https://www.alignmentforum.org/posts/ZwshvqiqCvXPsZEct/the-learning-theoretic-agenda-status-2023}

Li, L., Xie, T., \& Li, B. (2023, May). \textit{Sok: Certified robustness for deep neural networks.} In 2023 IEEE symposium on security and privacy (SP) (pp. 1289-1310). IEEE. \url{https://doi.org/10.48550/arXiv.2009.04131}

Loizos, C. (2023). \textit{StrictlyVC in conversation with Sam Altman, part two (OpenAI)} [Video] \url{https://youtu.be/ebjkD1Om4uw?si=vqqJNIw0ue81ruaa&t=1340}

Miotti, A., \& Wasil, A. (2023). \textit{Taking control: Policies to address extinction risks from advanced AI.} \url{https://doi.org/10.48550/arXiv.2310.20563}

MIT Technology Review. (2023). \textit{Video: Geoffrey Hinton talks about the “existential threat” of AI.} \url{https://www.technologyreview.com/2023/05/03/1072589/video-geoffrey-h inton-google-ai-risk-ethics/}

National Institute of Standards and Technology (NIST). (2023a). \textit{Artificial Intelligence Risk Management Framework (AI RMF 1.0)} \url{https://nvlpubs.nist.gov/nistpubs/ai/NIST.AI.100-1.pdf}

National Institute of Standards and Technology (NIST). (2023b). \textit{U.S. Artificial Intelligence Safety Institute} \url{https://www.nist.gov/artificial-intelligence/artificial-intelligence-safety-institute}

Nassim, N. T. (2007). The black swan: the impact of the highly improbable. NY: Random House.

Nuclear Regulatory Commission. (1983). \textit{Safety Goals for Nuclear Power Plant Operation} \url{https://www.nrc.gov/docs/ML0717/ML071770230.pdf}

OpenAI. (2023a). \textit{GPT-4 system card.} \url{https://cdn.openai.com/papers/gpt-4-system-card.pdf}

OpenAI. (2023b). \textit{Weak-to-strong generalization.} \url{https://openai.com/research/weak-to-strong-generalization}

OpenAI. (2024). \textit{Preparedness Framework (Beta).} \url{https://cdn.openai.com/openai-preparedness-framework-beta.pdf}

Oversight of A.I.: Principles for Regulation: Hearing before the Judiciary Committee Subcommittee on Privacy, Technology, and the Law, U.S. Senate, 118th Congr. (2023). (testimony of Dario Amodei). \url{https://www.judiciary.senate.gov/imo/media/doc/2023-07-26_-_testimony_-_amodei.pdf}

Park, P.S., Goldstein, S., O'Gara, A., Chen, M. \& Hendrycks, D. (2023). \textit{AI Deception: A Survey of Examples, Risks, and Potential Solutions.} \url{https://doi.org/10.48550/arXiv.2308.14752}

Perrow, C. (1999). \textit{Normal accidents: Living with high risk technologies.} Princeton university press.

Prime Minister’s Office. (2023, November 1). \textit{The Bletchley Declaration by Countries Attending the AI Safety Summit, 1-2 November 2023} \url{https://www.gov.uk/government/publications/ai-safety-summit-2023-the-bletchley-declaration/the-bletchley-declaration-by-countries-attending-the-ai-safety-summit-1-2-november-2023}

Tegmark, M., \& Omohundro, S. (2023). \textit{Provably safe systems: the only path to controllable AGI.} \url{https://doi.org/10.48550/arXiv.2309.01933}

The Logan Bartlett Show. (2023). \textit{Anthropic CEO on Leaving OpenAI and Predictions for Future of AI} [Video]. YouTube. \url{https://www.youtube.com/watch?v=gAaCqj6j5sQ&t=5885}

UK Government. (2023). The Bletchley Declaration by countries attending the AI Safety Summit. GOV.UK. \url{https://www.gov.uk/government/publications/ai-safety-summit-2023-the-bletchley-declaration/the-bletchley-declaration-by-countries-attending-the-ai-safety-summit-1-2-november-2023}

Wang, K., Variengien, A., Conmy, A., Shlegeris, B. \& Steinhardt, J. (2022). \textit{Interpretability in the Wild: a Circuit for Indirect Object Identification in GPT-2 small} \url{https://doi.org/10.48550/arXiv.2211.00593}

Wasil, A. (2023) \textit{Addressing Global Security Risks From Advanced AI.}  \url{https://medium.com/fidutam/addressing-global-security-risks-from-advanced-ai-e81cc54d0c90}

World Nuclear Association. (2022). \textit{Safety of Nuclear Power Reactors} \url{https://www.world-nuclear.org/information-library/safety-and-security/safety-of-plants/safety-of-nuclear-power-reactors.aspx}

Zou, A., Phan, L., Chen, S., Campbell, J., Guo, P., Ren, R., ... \& Hendrycks, D. (2023). \textit{Representation engineering: A top-down approach to AI transparency.} \url{https://doi.org/10.48550/arXiv.2310.01405}

\end{document}